\documentclass[10pt]{elsart}
\usepackage{epsfig,natbib}
\journal{New Astronomy}

\begin{document}
\runauthor{Bagla and Ray}
\begin{frontmatter}
\title{Performance Characteristics of TreePM codes}
\author{J.S. Bagla},
\ead{jasjeet@mri.ernet.in}
\author{Suryadeep Ray}
\ead{surya@mri.ernet.in}
\address{Harish-Chandra Research Institute, Chhatnag Road, Jhunsi,
Allahabad-211019}

\begin{abstract}
We present a detailed analysis of the error budget for the TreePM
method for doing cosmological N-Body simulations.  
It is shown that the choice of filter for splitting the inverse square
force into short and long range components suggested in Bagla (2002)
is close to optimum. 
We show that the error in the long range component of the force
contributes very little to the total error in force.  
Errors introduced by the tree approximation for the short range force
are different from those for the inverse square force, and these
errors dominate the total error in force.  
We calculate the distribution function for error in force for
clustered and unclustered particle distributions.  
This gives an idea of the error in realistic situations for different
choices of parameters of the TreePM algorithm.
We test the code by simulating a few power law models and checking for
scale invariance.
\end{abstract}


\end{frontmatter}

\section{Introduction}

Cosmological N-Body simulations have played a crucial role in
improving our understanding of formation of large scale structure.
These have filled large gaps in a domain where analytical solutions do
not exist.  N-Body simulations have also played a useful role by
testing scaling relations derived from physically motivated ansatze.

Limitations of computing resources and our ability to simulate
physical processes numerically have meant that N-Body simulations give
only approximate solutions.  It is possible to test whether
the approximations in evolution of the system affects
estimation of physical quantities of interest so the results are
generally more reliable than those based on approximate evolution of
the system.

A large number of methods have been used for doing Cosmological N-Body
simulations \citep{edbert_araa}.  All of these use some
approximation for calculation of  
force.  Approximations are used because direct summation over all
pairs of particles scales as $O(N^2)$, where $N$ is the number of
particles, making the calculation very time consuming for large $N$.
The use of these approximations reduces the number of calculations
required to $O(N\ln{N})$ or less.  These approximations also
introduce inaccuracies in the computed force.

The TreePM code \citep{treepm}, that we study here, is a hybrid technique
for carrying out large N-Body simulations to study formation and
evolution of large scale structures in the universe. 
It is a combination of the \citet{bh86} Tree code and a
Particle-Mesh code \citep{sim_book}.  
The TreePM method combines high
resolution of tree codes with the ability of PM codes to compute the
long range force with periodic boundary conditions.

In this paper we carry out a comprehensive study of the TreePM code.
We analyse errors in estimation of force in both the tree and the PM
components, and also study the distribution of errors for different
distributions of particles.  We also study the variation of CPU time
required for the TreePM code as we vary parameters describing the
mathematical model of this method.

\section{TreePM method}

A large number of methods have been suggested for improving the PM
method by combining it with other methods of computing force at small
separations \citep{pppm,cman,tpm} to reduce the softening length below
the grid 
length cutoff imposed by the PM method.  A common failing of nearly
all such methods is that these continue to use the usual PM force,
which is known to have large errors and anisotropies at scales
comparable to the grid scale
\citep{bouchet_pm}.  These errors are present in the final force, and
there is no natural way of reducing the errors in any of these
methods. 

The philosophy of TreePM method is to modify the PM force in
order to have a better control over errors in the long range force.
This is done by an explicit division of the potential and force into a
long range and a short range component.  PM method is then used to
compute only the long range component, and the tree method is used to
calculate the short range component.  This can be expressed in terms
of equations as:
\begin{eqnarray}
\varphi_{\bf k} &=& \varphi_{\bf k}^s + \varphi_{\bf k}^l \\
\varphi_{\bf k}^l &=& \varphi_{\bf k} F(k r_s,\ldots) \\
\varphi_{\bf k}^s &=& \varphi_{\bf k} \left(1 - F(k r_s,\ldots)\right)
\end{eqnarray}
Here, $\varphi_{\bf k}$ is the Fourier transform of the gravitational
potential, $\varphi_{\bf k}^s$ and $\varphi_{\bf k}^l$ are the short
range and the long range components, respectively.  $F(k r_s,\ldots)$
is a filter function that is unity for $k r_s \ll 1$ but decreases
rapidly for $k r_s > 1$.  Thus $r_s$ is the scale at which the
short range and long range forces are split.  $F$ can depend on other
parameters apart from $r_s$.  

This manner of splitting force permits us to control errors in both
the components of force.  What we require is a function $F(k
r_s,\ldots)$ that decreases sufficiently quickly at $k r_s > 1$.
Similarly in real space, the short range potential and force should
become negligible beyond a few $r_s$, and the long range force should
be negligible below $r_s$.  \citet{treepm} suggested $F =
\exp\left[-k^2r_s^2\right]$.  This function has an added feature that
it is positive definite, as is its Fourier transform.  Relevant
equations for this case are:
\begin{eqnarray}
\varphi_{\bf k}^l &=& - \frac{4 \pi G \varrho_{\bf k}}{k^2}
\exp\left[-k^2r_s^2\right] \label{pmf} \\
\varphi^l(r) &=& - \frac{Gm}{r} {\rm erf}\left(\frac{r}{2 r_s}\right) \\
\varphi^s(r) &=& - \frac{Gm}{r} {\rm erfc}\left(\frac{r}{2 r_s}\right) \\
{\bf f}^l({\bf r}) &=& - \frac{G m {\bf r}}{r^3} \left({\rm
erf}\left(\frac{r}{2 r_s}\right) - \frac{r}{r_s \sqrt{\pi}}
\exp\left(-\frac{r^2}{4 r_s^2}\right)\right) \label{flong} \\
{\bf f}^s({\bf r}) &=& - \frac{G m {\bf r}}{r^3} \left({\rm
erfc}\left(\frac{r}{2 r_s}\right) + \frac{r}{r_s \sqrt{\pi}}
\exp\left(-\frac{r^2}{4 r_s^2}\right)\right) \label{fshort}
\end{eqnarray}
Here, ${\rm erf}$ is the error function and ${\rm erfc}$ is the
complementary error function.  Most of our study is restricted to the
TreePM codes that use this filter.

\begin{figure}
\epsfysize=2.55truein\epsfbox[38 23 519 505]{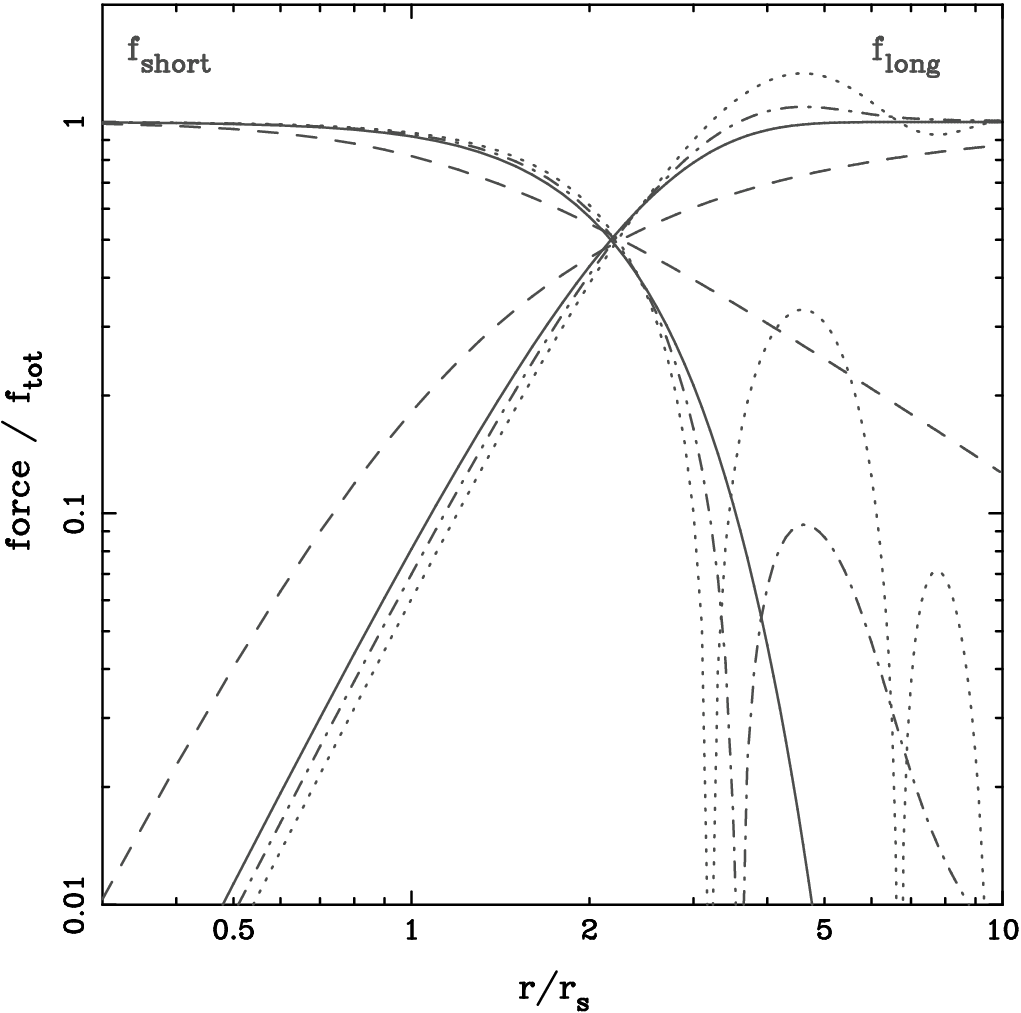}
\epsfysize=2.55truein\epsfbox[10 23 519 505]{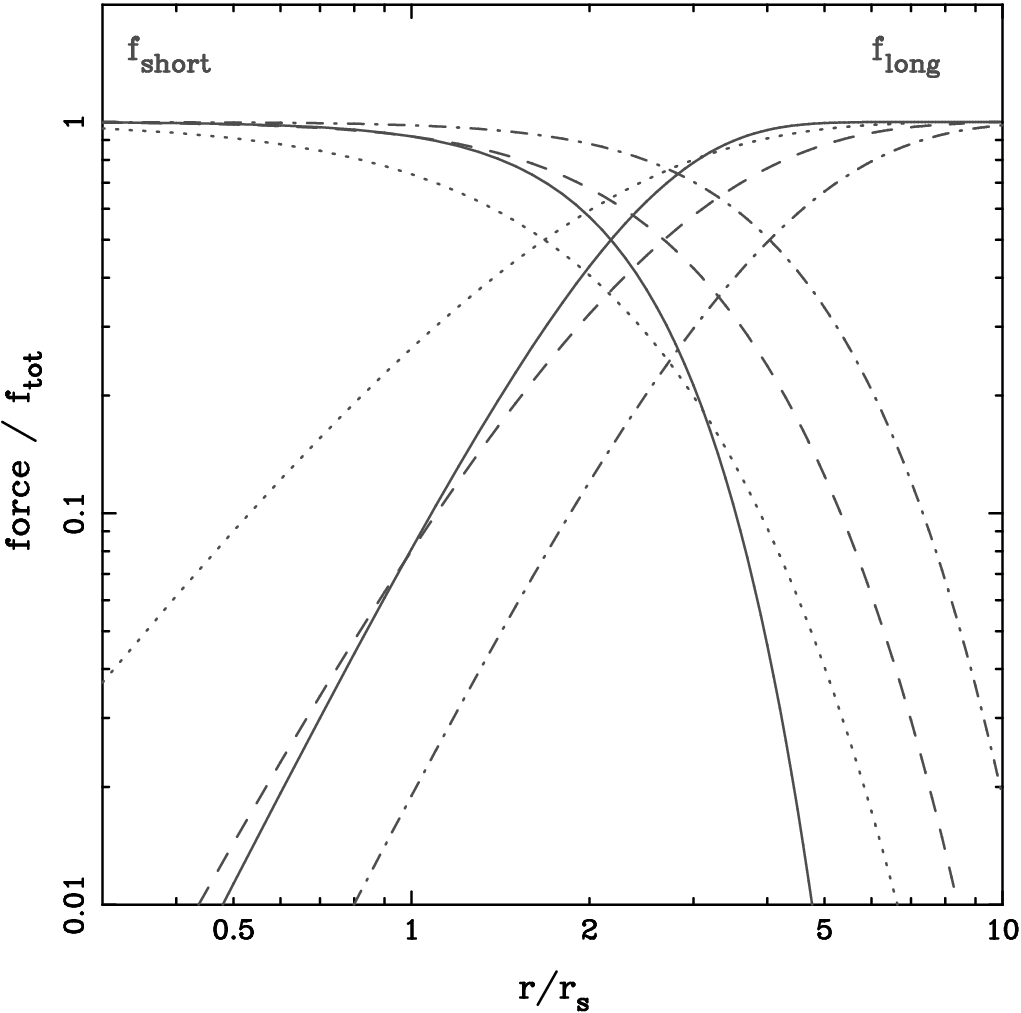}
\caption{We have shown the variation of the short range and the long
range forces for a choice of filters in this figure.  We have plotted
the ratio of short-range/long-range force and the inverse square force
as a function of distance scaled by the scale used for partitioning.
Left panel shows 
curves for filters described by eqn.({\ref{almods}}), and the right
panel shows curves for filters described by eqn.({\ref{nmods}}).  For
reference, we 
have also plotted the curve corresponding to $\alpha=2$ on the
left panel.  In the left panel, dashed curve, solid curve, dot-dashed
curve and dotted curve correspond to $\alpha = 1.0,2.0,2.5$ and $4.0$
respectively.  In the right panel, the solid curve refers to $\alpha =
2.0$ and the dotted curve, dashed curve, and dot-dashed curve
correspond to $n = 1,2$ and $4$ respectively.}
\end{figure}

We can generalise to more complex filter functions.  We studied two
families of functions:
\begin{eqnarray}
F(k r_s, \alpha) &=&
\exp\left[-\left(k^2r_s^2\right)^{\alpha/2}\right] \label{almods}\\
F(k r_s, n) &=& \left( 1 + k^2 r_s^2 \right)^{-n} \label{nmods}
\end{eqnarray}
Here $\alpha$ and $n$ are additional parameters.  For $\alpha=2$ in
eqn.(\ref{almods}), we recover the one 
parameter filter used in \citet{treepm}.  Fig.1 shows the ratios
$|f^s/f^{tot}|$ and $|f^l/f^{tot}|$ as functions of scale $r$ for these
models.  We have chosen $r_s=1$ for this figure.  Left panel shows
these ratios for models described by eqn.(\ref{almods}) and the right
panel shows models described by eqn.(\ref{nmods}).  In the left panel,
we see that the short range force for $\alpha < 2$ decreases very
slowly at large scales, making it an unsuitable choice.  For $\alpha >
2$, the short range force oscillates at large scales and the amplitude
of peaks decreases very slowly with scale.  Hence these models too are
not useful as an alternative to the $\alpha=2$ model.  Models
described by eqn.(\ref{nmods}) are shown in the right panel.  In this
case the behaviour is better than models with $\alpha \neq 2$ in that
the long and the short range forces both fall off much faster.
However, the model with $\alpha = 2$ is better than any of these
models.  This model is plotted in the right panel as the solid curve.
We would like the overlap between the short and long range forces to
be as small as possible.  For one, a rapidly decreasing short range
force implies that we need to sum this over a small region in space.
Computation of the short range force is the most time consuming part
and hence a sharply falling short range force is required for a well
optimised code.  The second reason which requires us to have a long
range force that decreases rapidly below some scale is that the PM
force typically has large errors at around grid scale.  The main
motivation of considering an explicit division of force in a long
range and a short range component was to avoid carrying over the
errors in the PM force by truncating the long range force at scales
where PM contributes most significantly to errors.  Thus we want to
use that filter for which the range of scales at which both the long
range and the short range components are relevant is the smallest, and
the Gaussian filter is the optimum one in this regard and no other
filter comes close to this.  
We shall not discuss any other filter from this point onwards as
figure~1 suggests that the Gaussian filter is the most optimum one.

\section{Error Analysis}

We shall first study errors in the force field of one particle.  As
the tree approximation does not have any errors for a single particle,
we shall focus exclusively on errors in the long range force.

\subsection{Errors in the Long Range Force}

The long range force (eqn.(\ref{flong})) is computed by the
Particle-Mesh method.  We summarise the PM method here, and for
details we refer the reader to other sources
\citep{sim_book,bouchet_pm,nbody_pm}.  Particle positions are used to
compute 
the density field on a regular mesh.\footnote{We will use terms mesh
and grid interchangeably in the context of PM codes.}  This is done
using an interpolating function $W(x,x_i)$ that assigns mass of a
particle at position $x$ to mesh points $x_i$.  In three dimensions,
the interpolating 
function is a product of three one dimensional interpolating
functions.  Density field on the mesh is Fourier transformed using
Fast Fourier transform (FFT).  In Fourier space, we solve
eqn.(\ref{pmf}) instead of the usual Poisson equation.  We multiply
this with $\iota {\bf k}$ to get the negative gradient of the long
range force in Fourier space.  An inverse transform using FFT gives us
the long range force on mesh points.  This is interpolated back to
particle positions using the same function $W$.  In this scheme, 
mesh and the interpolating function $W$ are the main sources of
anisotropy.  We use the cloud-in-cell (CIC) interpolating
function~\citep{sim_book}.  
\begin{eqnarray}
W(x,x_i) &=& 1 - {|x-x_i|}/L ~~~~ |x-x_i| \leq L \\
&=& 0 {~~~~~}{~~~~}{~~~}{~~~}{~~~} |x-x_i| > L \label{cic}
\end{eqnarray}
Here $L$ is the separation between adjacent mesh points.

To estimate errors in the long range force, we follow the method
described by \citet{pppm}.  We place a particle at a random position
in a mesh cell and use our code to find the long range force.  The
force is evaluated at a large number of random points scattered within
some distance from the particle.  This process is repeated a number of
times with a different position of the particle.  We use the force
calculated this way to compute the average long range force as a
function of distance as well as the dispersion about this average.

\begin{figure}
\epsfxsize=5.4truein\epsfbox[40 23 519 505]{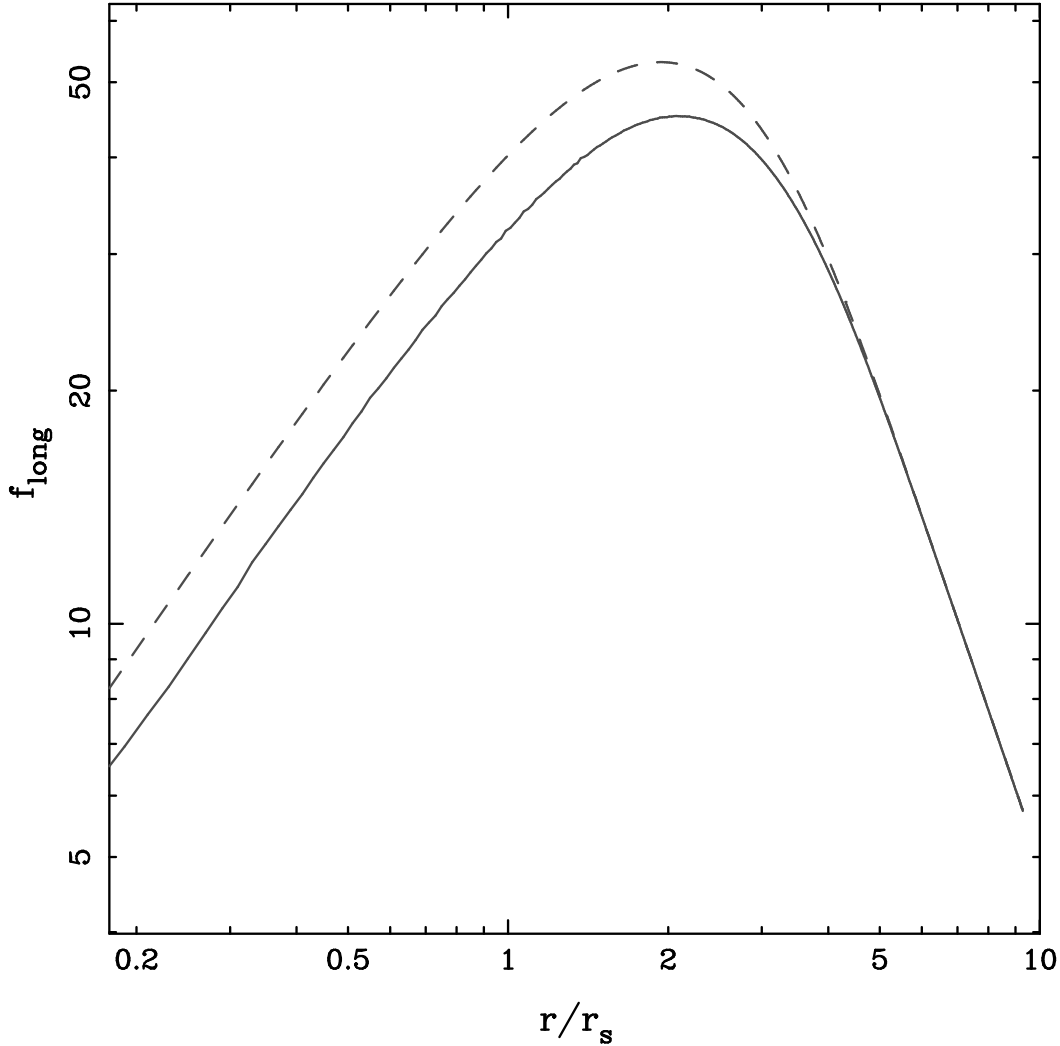}
\caption{Long range force for a single particle is shown as a function
of distance.  The dashed curve shows the expected long range force
(eqn.({\ref{flong}})) and the thick curve shows the long range force
obtained in the simulation.  This curve was drawn for $r_s=L$.  There
is a clear under-estimation of the long range force at small scales.
See text for details.} 
\label{flong_r}
\end{figure}

Figure~\ref{flong_r} shows a plot of the average long range force
(solid curve) and 
eqn.(\ref{flong}) (dashed curve) as functions of scale.  This curve was
drawn for $r_s=1$.  The average force clearly falls below the expected
values at small scales.

\begin{figure}
\epsfysize=2.55truein\epsfbox[38 28 515 507]{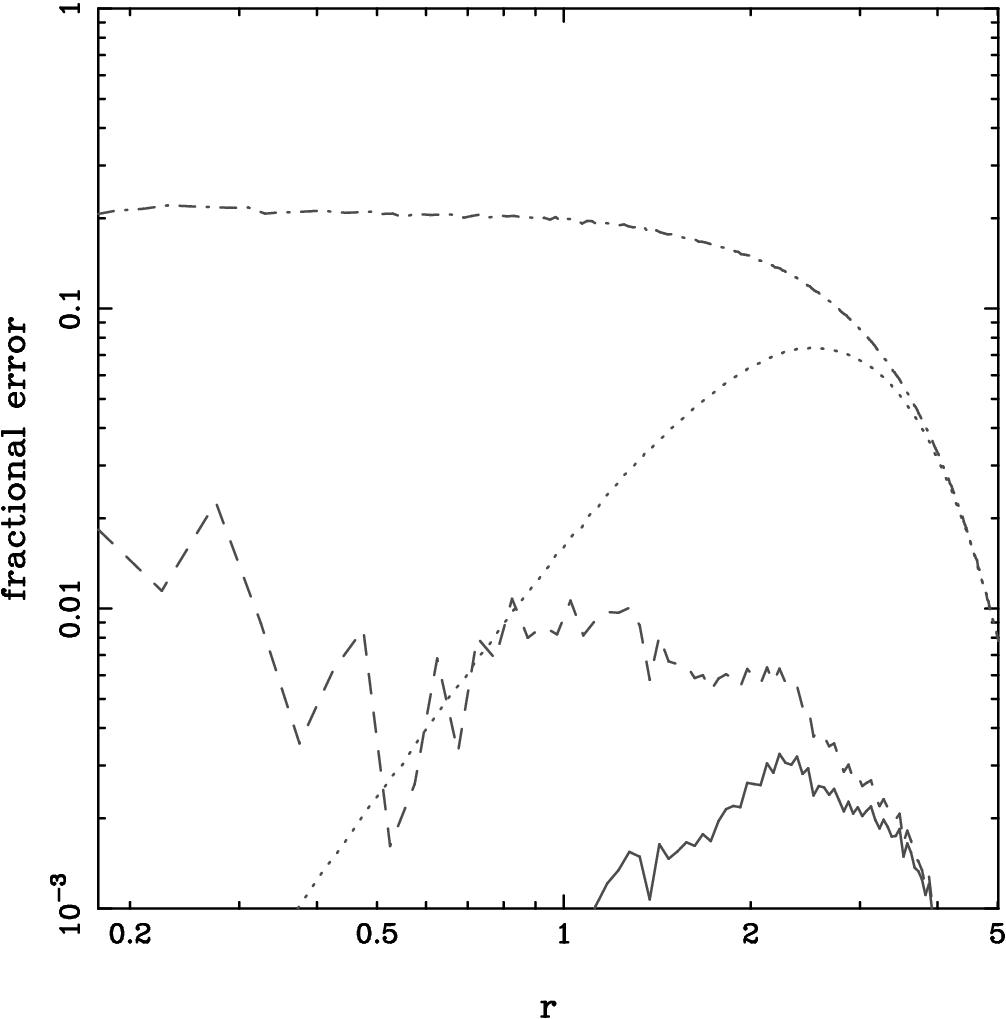}
\epsfysize=2.55truein\epsfbox[10 23 512 505]{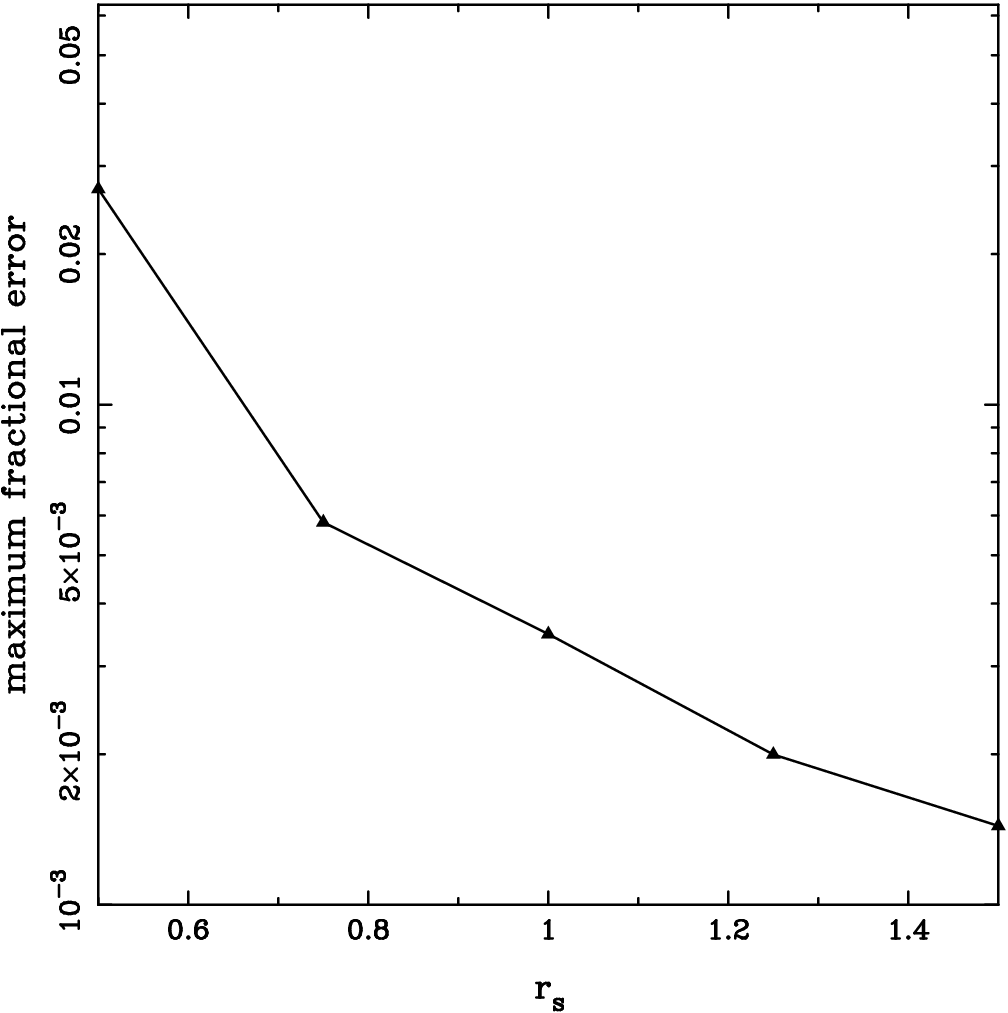}
\caption{Left panel shows the fractional error in the long range force
due to a particle as a function of scale.  Difference of the computed
long-range force from eqn.({\ref{flong}}) is scaled by the expected long
range force for the dot-dashed curve.  The same difference is scaled
by the inverse square force and shown by the dotted curve.  This
latter curve is more relevant for studying the contribution to the
error in force.  Curves showing the fractional error after removing
the effect of interpolating function are also shown here.  Dashed
curve shows the fractional error when scaled with the long range force
and the solid curve shows the fractional error scaled with the inverse
square force.  See text for details of how the effect of interpolating
function is removed.  The peak fractional error when scaled by the
inverse square force for this case is less than $0.4\%$ for $r_s=L$.
The right panel shows the peak fractional error as a function of
$r_s$.}
\label{flongerr_r}
\end{figure}

This difference is caused mainly by the interpolating function.
In the continuum limit we require the interpolating function to be
like the Dirac delta function.  However, we cannot use that and hence
there is a serious shortfall in the force at small scales.  This
becomes clear in Fourier space analysis \citep{bouchet_pm,sim_book}.
Same analysis also suggests that by de-convolving the interpolating
function, we should be able to recover the expected long range force.
We deconvolve\footnote{We divide the potential in k-space by the
square of the Fourier transform of the interpolating function.  Since
the three dimensional interpolating function is product of the three
one dimensional interpolating functions, we are essentially dividing
$\varphi_k$ by $W_{k_x}^2 W_{k_y}^2 W_{k_z}^2$.  Square of each of
these needs to be taken as we need to do the deconvolution two times.}
the interpolating 
function twice, as we use it twice in 
force calculation: once for calculating the density on the mesh and
then for calculating force at the positions of particles.  For
reference, we mention that the Fourier transform of the one
dimensional CIC function is $4\sin^2(k L/2) / (k L)^2$, where $k$ is
the wave number and $L$ is the size of one mesh cell.

Such deconvolution of an arbitrary force will not work very well as it
will only enhance anisotropies introduced by the mesh.  However in
this case contribution of wave modes near the Nyquist frequency is
negligible due to the Gaussian filter.  So we are
essentially trying to correct for the amplitude at intermediate wave
numbers.  Partial correction can be obtained by de-convolving for the
effect at small wave numbers by considering the Taylor expansion of
$W_k$ about $k=0$ \citep{mw_prv}.

\begin{figure}
\epsfysize=2.55truein\epsfbox[40 28 519 505]{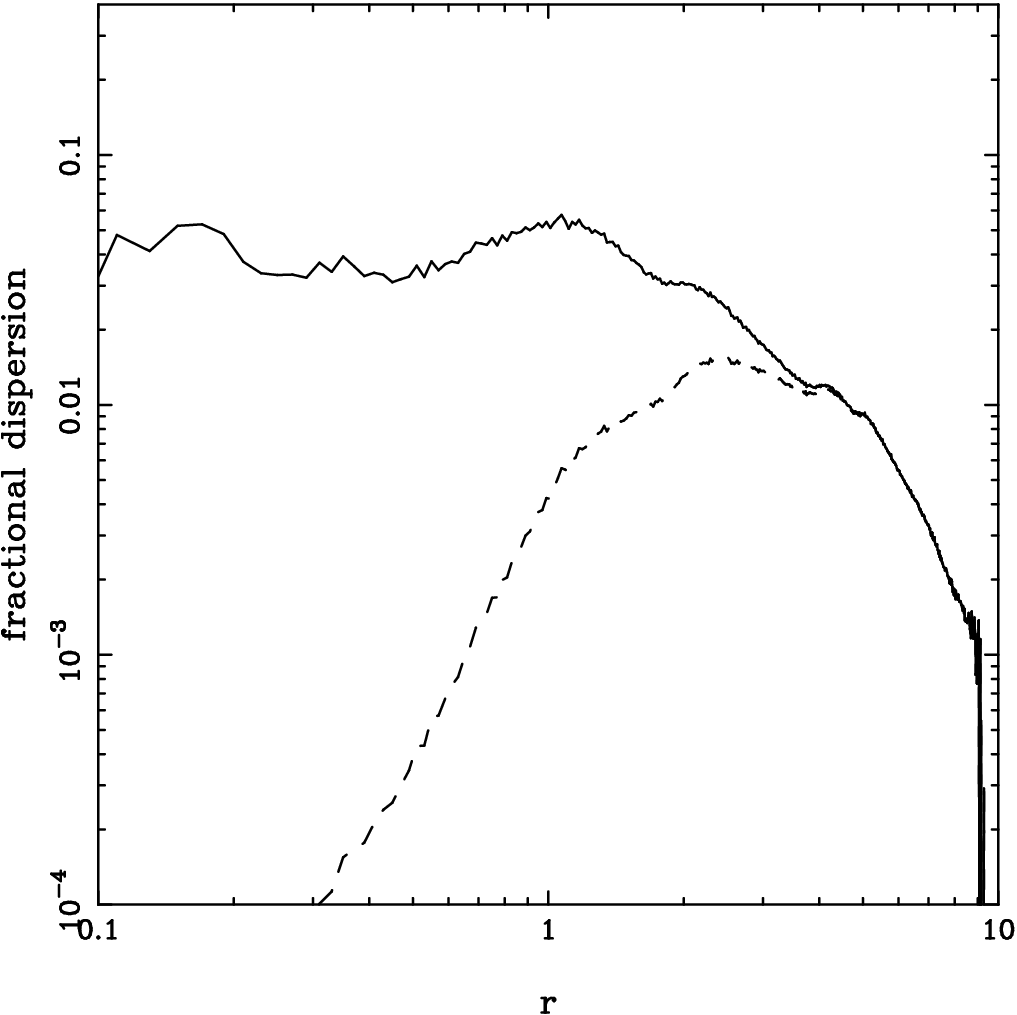}
\epsfysize=2.55truein\epsfbox[10 23 512 505]{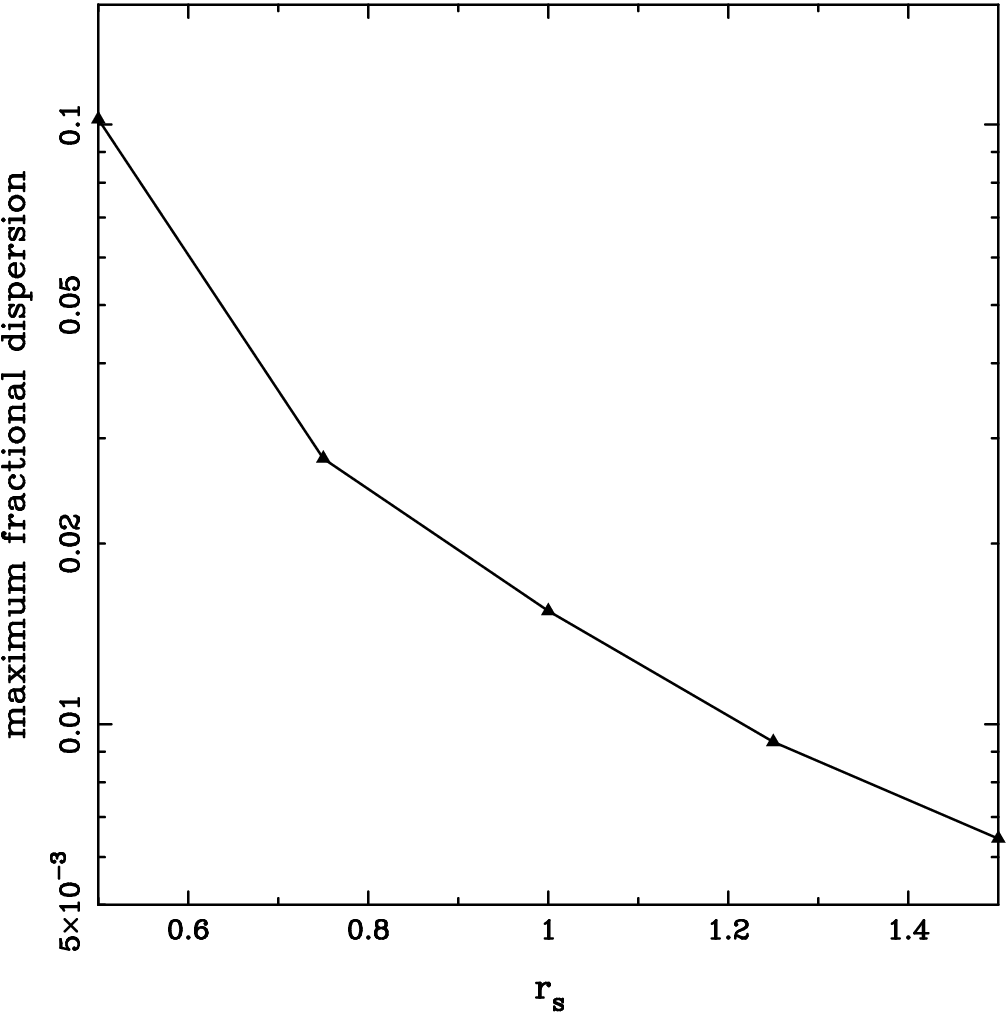}
\caption{Dispersion about the average long range force for $r_s=1$
with deconvolution of the interpolating window function.  This
dispersion scaled by the long range force is shown by the solid
curve.  Dispersion scaled by the inverse square force is shown by the
dashed line.  The peak dispersion in this case is just over $1.4\%$ of
the inverse square force. The right panel shows the maximum value of
dispersion scaled by the inverse square force as a function of $r_s$.}
\label{disp_flong_rs}
\end{figure}

The effect of deconvolution on the long range force is shown in
figure~\ref{flongerr_r}.  We have plotted the difference of the
average long range force and eqn.(\ref{flong}) divided by the long
range force.  We have also plotted the difference divided by the total
force because it is this quantity that is relevant for errors in the
total force.  We have plotted these for the average force shown in
figure~\ref{flong_r} as well as the average force obtained after
deconvolution described above.  The effect of deconvolution is
striking---the 
difference between the average long range force and eqn.(\ref{flong})
normalised by the total force drops by nearly one order of magnitude.
Fractional error drops below $1\%$ and is below $0.4\%$ for the
entire range of scales.  One can lower this further by using a larger
$r_s$, or use a lower $r_s$ if larger errors are acceptable.
Figure~\ref{flongerr_r} also shows the maximum of the difference between
the average long range force and eqn.(\ref{flong}) normalised by the
inverse square force as a function of $r_s$.  As expected, this
decreases as we 
increase $r_s$.  Thus we can choose $r_s$ once we have fixed the error
that is acceptable for our physical application.  We will use long
range force with deconvolution of the interpolating function $W$ in
the following discussion.

From deviations of the average force from eqn.(\ref{flong}), we now
turn to the dispersion in force about this average.  Both these
quantities are calculated in narrow bins in distance $r$.
Figure~\ref{disp_flong_rs} shows the dispersion in the long range
force.  We have plotted {\it rms} dispersion about the average force
as a function of $r$.  We have normalised the dispersion with the long
range force (eqn.(\ref{flong})), and the total force.  The latter
curve is relevant for total force.  From this curve we
can see that the rms dispersion about the average long range force
reaches a peak of about $1.4\%$ near $r=2 r_s$.  

The dispersion, just like the deviation of the average long range
force from the theoretical expectation eqn.(\ref{flong}), decreases as
we increase $r_s$.  Figure~\ref{disp_flong_rs} shows the variation of
the maximum of {\it rms} dispersion normalised by the inverse square
force, as a function of $r_s$.  

We can summarise our investigations of the errors in the long range
force for a particle as follows:  Maximum deviation of the long range
force from the theoretical expectation is around $0.4\%$ and this
deviation is much smaller at most scales.  Dispersion about the
average peaks at a value of $1.4\%$ and is well below $1\%$ for most
scales.  These values are for $r_s=L$ and the maxima of errors occur
around $2r_s$, errors fall off rapidly on both sides. 
Errors decrease as we increase $r_s$. 

We will not study errors in the long range force for a system of
particles.  We will, instead, study the error in the total force after
studying the errors in the short range force.

\section{Errors in the Short Range Force}

The short range force is calculated using the tree approximation.  We
use the \citet{bh86} method.  In this method, the simulation volume is
taken to be a cube.  Particles are arranged in a hierarchy of cells in
an oct-tree, i.e., each cell can have up to eight daughter entities.
These can either be cells or particles.  
The simulation volume is divided into smaller cells so that at the
smallest level, no cell contains more than one particle.  
Once the tree structure is set up, we calculate the position of the
centre of mass for each cell as well as the total mass contained
within it.

Tree approximation is then used to calculate force on each particle.
In this approximation, distant cells are treated as point masses with
all the mass concentrated at the centre of mass.  Error in force
introduced by this approximation is proportional to the square of the
ratio of size of the cell ($d$) to the distance ($r$) to the centre
of mass from the point where force is being calculated.  We define
this ratio as $\theta = d/r$.  The error in force is proportional to
$\theta^2$. 
In order to control errors, we would not like to use cells that are
too close or too large.
This is done by introducing a cell acceptance criterion, e.g. by
computing the quantity $\theta$ and comparing it with a threshold
$\theta_c$: 
\begin{equation}
\theta = \frac{d}{r} \leq \theta_c  
\label{trwalk}
\end{equation}
The error in force will obviously increase with $\theta_c$.  
It can be shown that the maximum error for inverse square force is
$9\theta_c^2/4$.
For a detailed discussion of cell acceptance criteria and error in
force, we refer the reader to several detailed studies
\citep{perf_tree,skel,gadget}.

The tree approximation is used here for computing the short range
force which differs from the usual inverse square force.  We expect errors
to be different in this case.  We calculate the error at leading order
in tree approximation and compare it with the corresponding error in
the case of inverse square force. 

We take a cubical cell of dimension $d$ and distribute point masses
randomly inside it. Our aim is to calculate the error involved 
in computing the short range force in the tree approximation on a unit
point mass situated at the origin due to the masses in the cubical cell
described above.  
To calculate the error, we need to compare the short range force in
the tree approximation (${{\bf f}^s_{cm}}$) with the sum of short
range forces due to all the particles in this cell (${\bf f}^s$).
We also have the inverse square force in the tree approximation ${{\bf
f}_{cm}}$. 

\begin{figure}
\epsfysize=2.6truein\epsfbox[42 23 512 505]{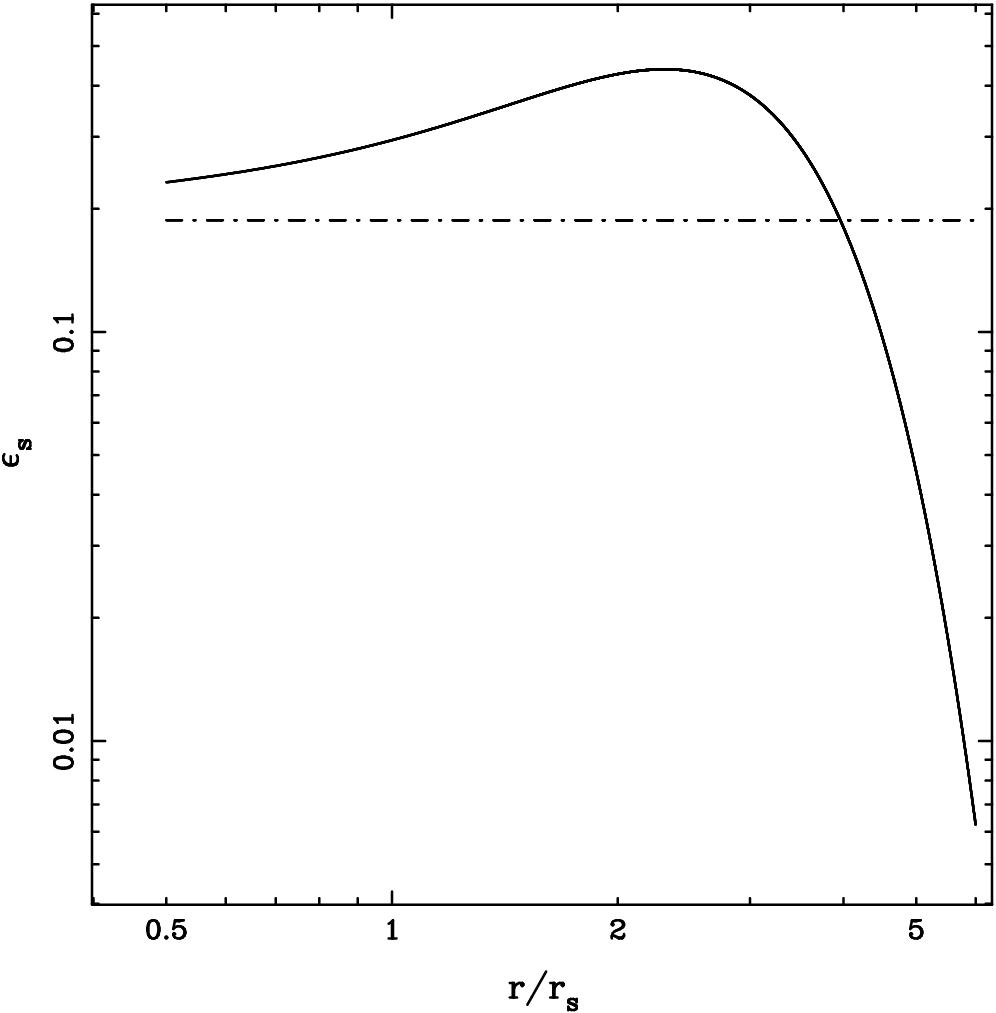}
\epsfysize=2.6truein\epsfbox[10 20 515 506]{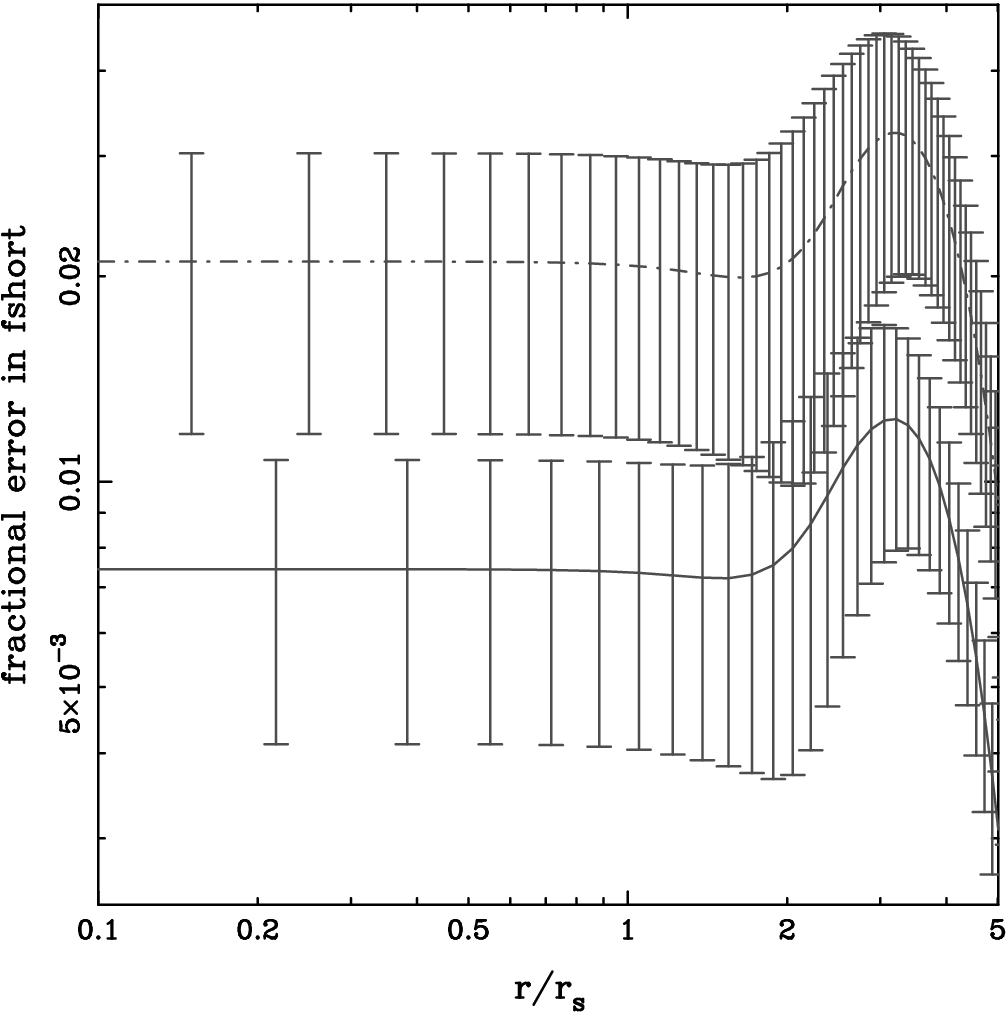}
\caption{Fractional error in the tree approximation in one of the
worst cases is shown in the
left panel as a function of distance $r$ for the short range force
(solid curve).  This is the error up to leading order in $\theta_c$
(see text).  For reference, we have also plotted the corresponding
error in case of the inverse square force (dot-dashed curve).  This
error is computed for the case when the two particles are separated
along an edge of the cell and the two particles and the point where
force is being calculated are along a line.  The right panel shows
fractional error in short range force measured using the 
method described in text. Here, the solid curve corresponds to
$\theta_c = 0.3$ and $N_p = 30$ and the dot-dashed curve is for
$\theta_c = 0.5$ with the same number of particles.  Error bars mark
$1\sigma$ variation about the average error measured for $100$
different distributions of particles.}
\label{fs_error}
\end{figure}

To quantify error in short range force we define ${\epsilon}^s$ as a
measure of the error in the computed short range force.
\begin{equation}
\epsilon_s = \frac{\left|{{\bf f}^s_{cm}} - {{\bf
f}^s}\right|}{|{{\bf f}_{cm}}|} 
\label{short-range-error}
\end{equation} 
We compute the error in case of the inverse square force for
comparison.  The worst case we can have is if two particles are at 
diagonally opposite ends of the cell, and the origin is along the same
diagonal.  The distance between the particles is $d\sqrt{3}$, and the 
fractional error in force upto leading order in $\theta_c$ is $9
d^2/(4 r^2)=9\theta_c^2/4$ where $r$ is the distance from the origin
to the centre of mass.   If the two particles are separated along an
edge of the cell instead of the diagonal, the error is
$3\theta_c^2/4$. 
A similar analysis for the short range force gives
\begin{eqnarray}
\epsilon_s &=& \frac{3 \theta_c^2}{4} \left[{\rm
erfc}\left(\frac{r}{2 r_s}\right)+ {\left(\frac{r}{r_s}\right)}
\exp\left(-\frac{r^2}{4 r_s^2}\right)\right] \nonumber \\ 
&& + \frac{\theta_c^2}{4} \left[\frac{2}{\sqrt\pi}  
{\left(\frac{r}{r_s}\right)^3} \exp\left(-\frac{r^2}{4 r_s^2}\right) +
\frac{1}{8 \sqrt\pi} {\left(\frac{r}{r_s}\right)^5} 
\exp\left(-\frac{r^2}{4 r_s^2}\right)\right].
\label{eps_fs}
\end{eqnarray}
Here we have also assumed that the separation between the two
particles is smaller than $r_s$.  Error here has been calculated for
the case when the two particles are separated along an edge of the
cell and the origin (where the force is being calculated) and these
two points are collinear.  We have plotted the error as a function of
distance $r$ in fig.{\ref{fs_error}}.

Now we turn to a more realistic estimate of error for random
distribution of particles in the cell, instead of considering
pathological situations which give rise to large error.
Right panel of figure~{\ref{fs_error}} shows the fractional error in
short range 
force $\epsilon_s$ as a function of distance $r$.  
Error is constant at small scales, increases around the scales where
the contribution of short range force is decreasing rapidly, then
plummets to very small values at larger scales.  
The range of scales where the error is large are those where the short
range force contributes a 
significant fraction to the total force {\sl and its variation with
distance is significantly different from inverse square.}
In the figure, error is plotted for two values of $\theta_c$.  
It is lower for smaller $\theta_c$: error for $\theta_c=0.3$ is a
nearly a factor three smaller than for $\theta_c=0.5$.  
Here error has been averaged over $100$ different distributions of
$30$ particles in the cell.
Error is smaller for larger number of particles as chances of an extreme
distribution of particles become smaller.
Even for $N_p=10$, the error is considerably smaller than the extreme
case we used for estimating errors above.
Error for $\theta_c=0.5$ and $N_p=30$ is about $2\%$, slightly larger
than the
maximum error due to long range force~($1.4\%$ for $r_s=L$). 
Errors for $\theta_c=0.3$ are smaller and closer to the errors in the
long range force.
However, we should add that these are errors due to particles in a
single cell.  
In a generic situation, these errors may cancel out or add up,
depending upon the details of the particle distribution.  

\section{Errors in the TreePM force}

We turn to the question of errors in the TreePM force calculated
by adding the short range and the long range forces.  
Errors are calculated with respect to a reference force computed with
the following configuration of the TreePM code.
We chose $\theta_c=0.01$, hence the errors in short range force are
negligible as these errors decrease in proportion to $\theta_c^2$ for
the worst case scenario.
We take $r_s=4.0$ thus error in the long range force is 
below $0.2\%$ if we extrapolate the maximum dispersion in long
range force.
With this setup, we can safely assume that the error in reference
force (as compared to the inverse square force with periodic boundary
conditions that we should use) is below $0.2\%$ at all scales,
and this is sufficient for studying errors in the TreePM force
calculated using more pragmatic values of $r_s$ and $\theta_c$ as we
are aiming to keep errors below $1\%$ level for most particles in all
situations. 
  
We will calculate errors for two very different distributions of
particles, one a random distribution of particles with uniform density
(unclustered distribution) and the other a clustered distribution
generated by an N-Body simulation.  
We will present the results for these distributions side by side.

We present the results of our analysis by plotting the cumulative
distribution for fractional error.  Errors were calculated for a
distribution of $2^{21} \simeq 2\times 10^6$ particles so that we have
sufficient number of particles to study the distribution. 

\begin{figure}
\epsfysize=2.6truein\epsfbox[40 25 521 507]{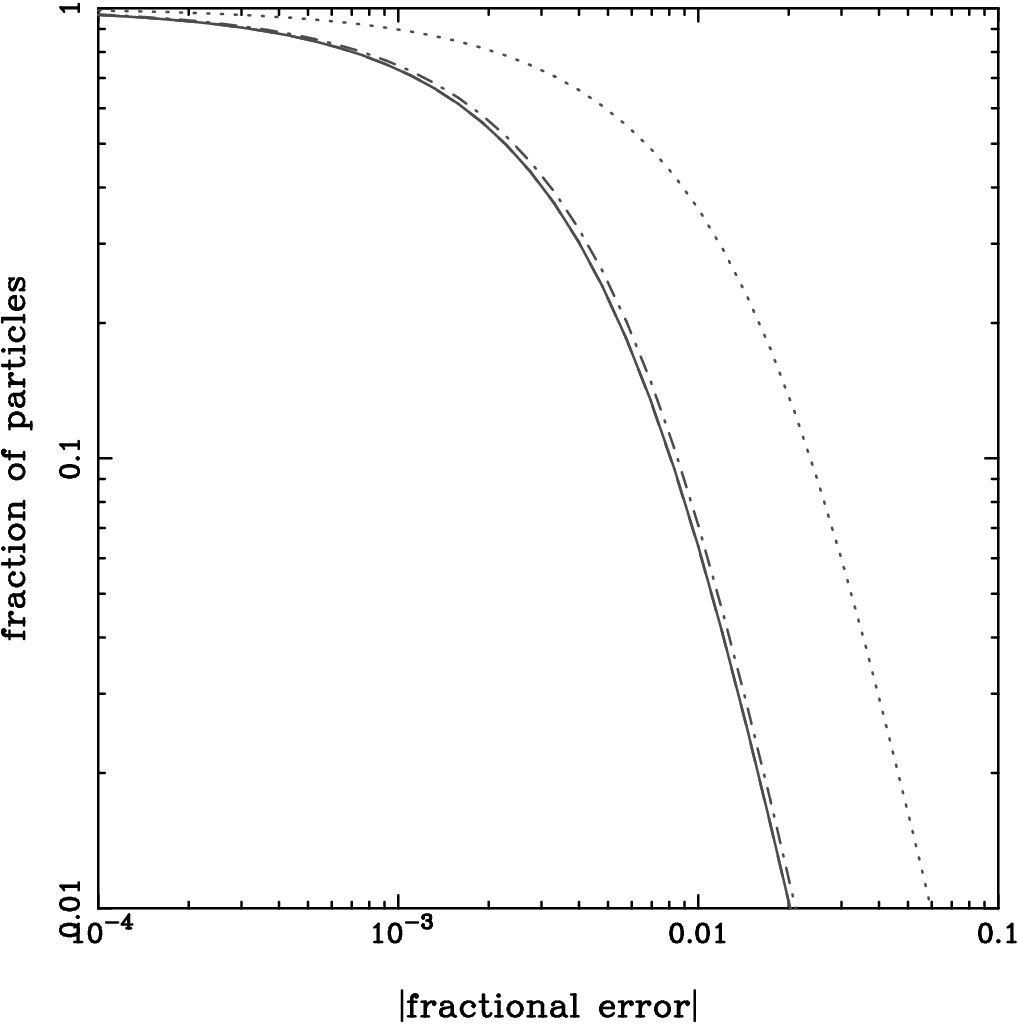}
\epsfysize=2.6truein\epsfbox[10 25 521 507]{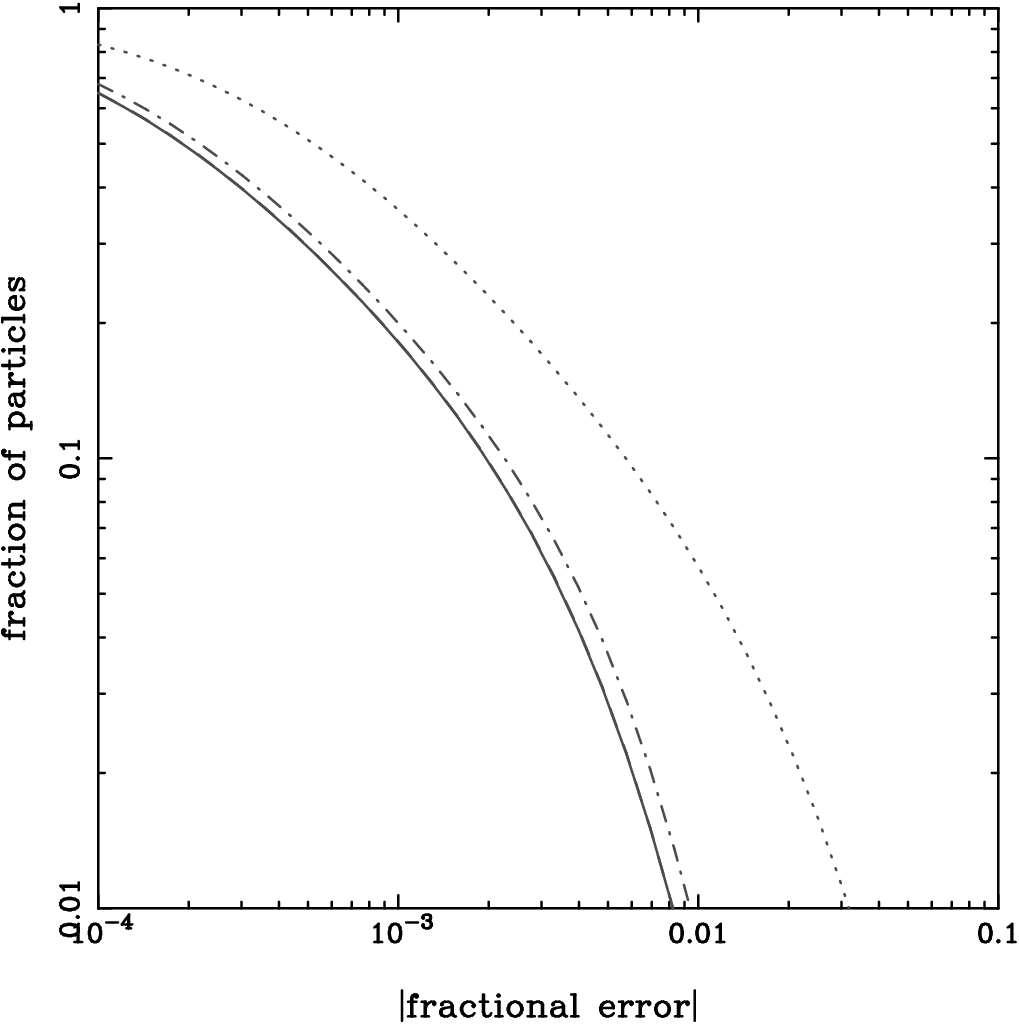}
\caption{The distribution of fractional error in the TreePM force is
shown here for two distributions of particles.  The left panel shows
the distribution for a uniform distribution of particles, for
different choices of $r_{cut}$, the radius within which the short
range force is summed.  The right panel shows the same for a clustered
distribution of particles.  We used $r_s=L$ and $\theta_c=0.5$ for
these figures. Solid curve corresponds to $r_{cut}=6 r_s$, dashed
curve is for $r_{cut}=5 r_s$, dot-dashed and dotted curves correspond
to $r_{cut} = 4 r_s$ and $3 r_s$ respectively.  The error is very 
large in both cases for $r_{cut}=3 r_s$.  The error for all
$r_{cut}\geq 4 r_s$ is essentially the same.  There is some difference
in the error for $r_{cut}= 4 r_s$ 
and $r_{cut}\geq 5 r_s$ for the clustered distribution, as
anisotropies in the particle distribution start to make the force from
particles at these distances more and more significant.}
\label{fig_rcut}
\end{figure}

We have presented the variation of errors with $r_{cut}$ in
fig.~{\ref{fig_rcut}}.  For this figure we used $r_s=L$ and
$\theta_c=0.5$. 
$r_{cut}$ is the radius within which we sum
the short range force around each particle.  We need to sum out to a
distance where the contribution of the short range force to the
inverse square force drops to an ignorable fraction.  From fig.~1,
we know that the short range force is less than $1\%$ of the inverse
square force at $5r_s$, so we expect that $r_{cut}$ should be
comparable to this value.  We find that for the unclustered
distribution, distribution of errors is very different for
$r_{cut}=3r_s$ as compared to $r_{cut} \geq 4r_s$.  For all $r_{cut}
\geq 4r_s$, distribution of errors is very similar.  This can
result if particles at scales beyond $4r_s$ are not
contributing to the force.  This is to be expected in a uniform
distribution of particles.  Distribution of errors for the clustered
distribution is shown in the right panel of fig.~{\ref{fig_rcut}}.
For the clustered distribution, the particle
distribution is not homogeneous and hence particles at larger
distances contribute significantly to the force on each particle.  As
this contribution increases, force from $r > 4r_s$ becomes more
relevant.  Error distribution for the clustered distribution shows
small differences between $r_{cut}=4r_s$ and $r_{cut} \geq 5r_s$.
From these figures we conclude that $r_{cut}=5r_s$ is a safe choice
and that $r_{cut}=4r_s$ will suffice for most purposes.

\begin{figure}
\epsfysize=2.6truein\epsfbox[40 25 521 507]{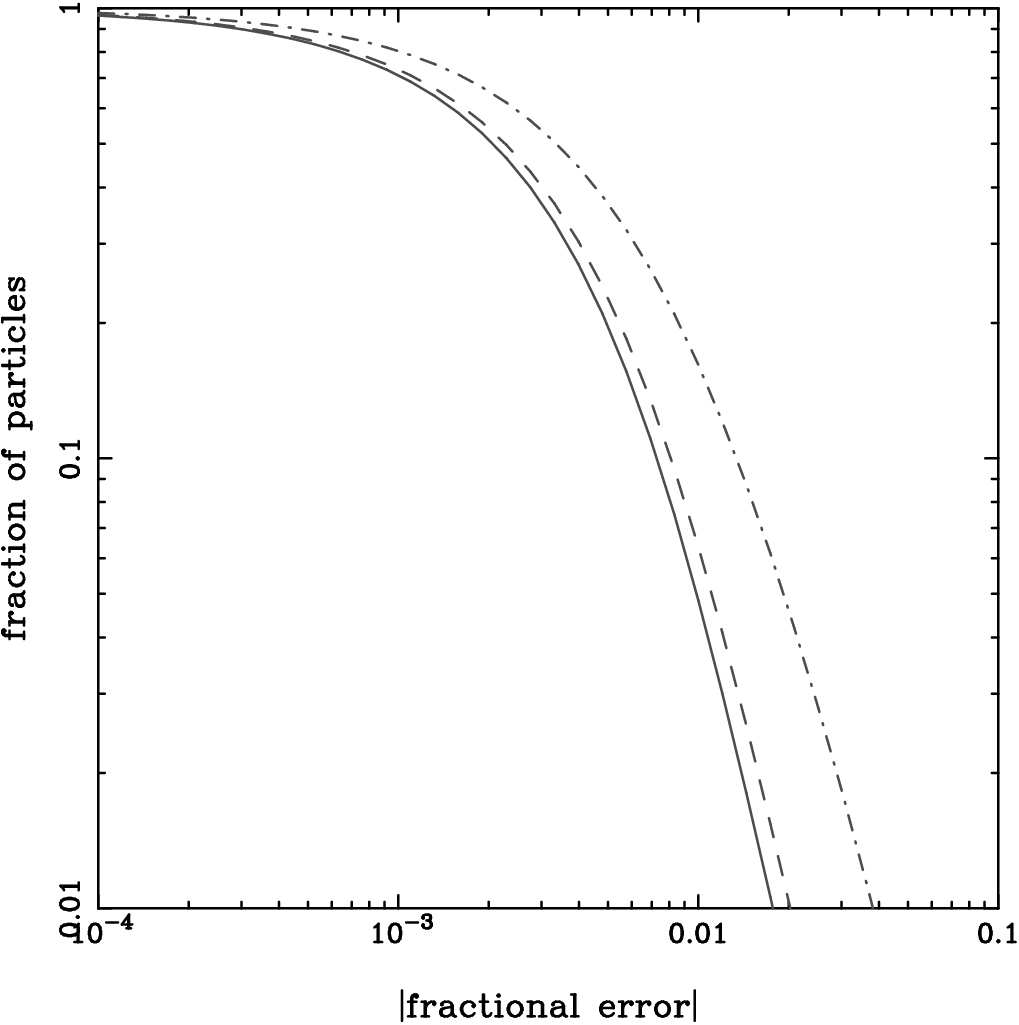}
\epsfysize=2.6truein\epsfbox[10 25 521 507]{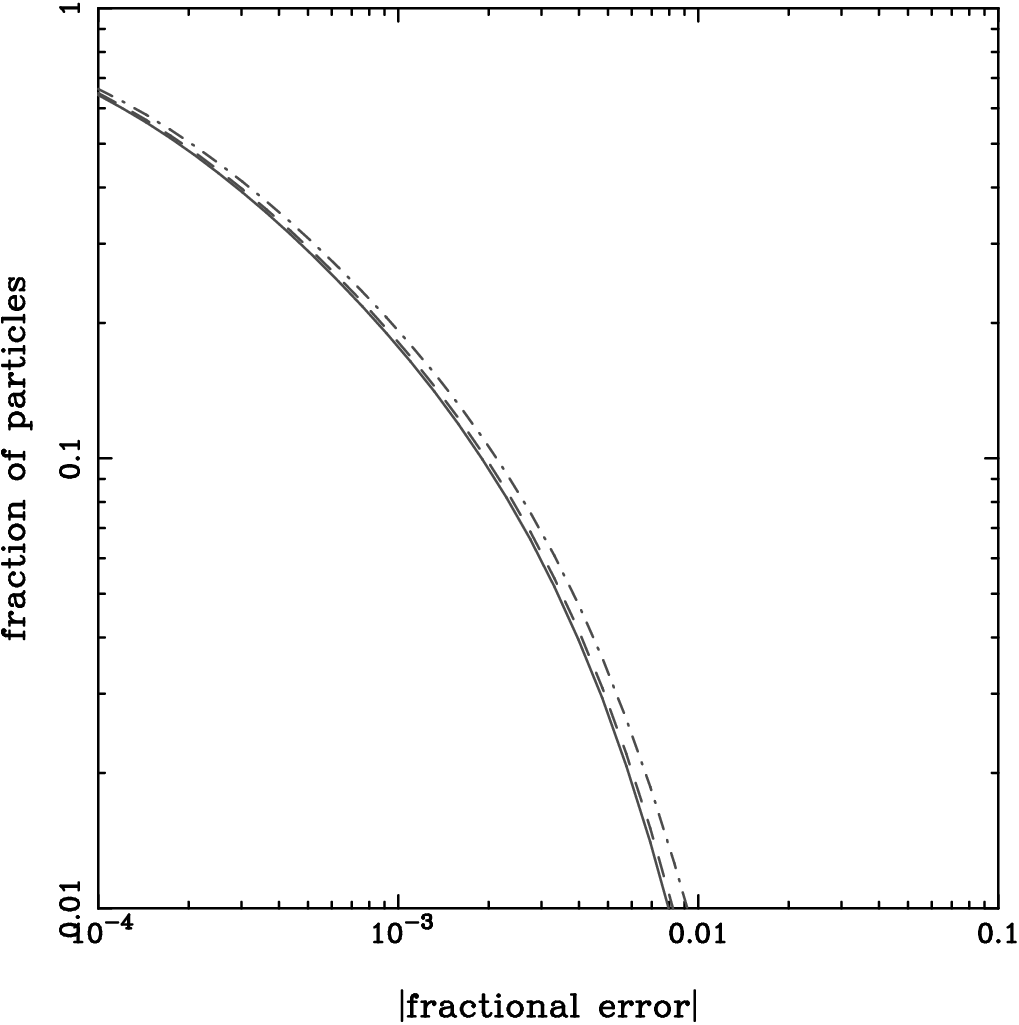}
\caption{The distribution of fractional error in TreePM force is shown
here for different values of $\theta_c$.  The left panel shows these
for a uniform distribution of particles and the right panel shows the
same for a clustered distribution of particles.  We used $r_s=L$ and
$r_{cut}=6r_s$ for these figures. Full line, dashed line and
dot-dashed line correspond to $\theta_c = 0.3, 0.5$ and 
$\theta_c = 0.7$ respectively in both panels. The order of curves
in both these cases is as expected, error is smaller for smaller
$\theta_c$.  The difference in error with different $\theta_c$ reduces
with increasing clustering.  See text for an explanation.}
\label{fig_thetac}
\end{figure}

Figure~{\ref{fig_thetac}} shows the distribution of errors for
different values of $\theta_c$.  As before, we have shown results for
the unclustered distribution in the left panel and that for the
clustered distribution in the right panel.  
We used $r_s=L$ and $r_{cut}=6r_s$ for these plots.
There is a marked difference in
errors for different values of $\theta_c$ in case of unclustered
distribution.  Error decreases with $\theta_c$, the variation
decreases as we get to smaller values of $\theta_c$.  Situation is
somewhat different for the clustered distribution as there is hardly
any difference between errors for different values of $\theta_c$.
This suggests that the errors are dominated by errors in the long
range force for clustered distribution.  For unclustered distribution,
the total force on each particle is small, whereas force due to a cell
with many particles is large and many large contributions cancel out
to give a small net force.  Numerical errors in adding and subtracting
large numbers seem to dominate errors in case of the uniform
distribution.  Contribution of cells is large for larger $\theta_c$,
hence we get a significant variation with $\theta_c$ for unclustered
distribution.

\begin{figure}
\epsfysize=2.6truein\epsfbox[40 25 521 507]{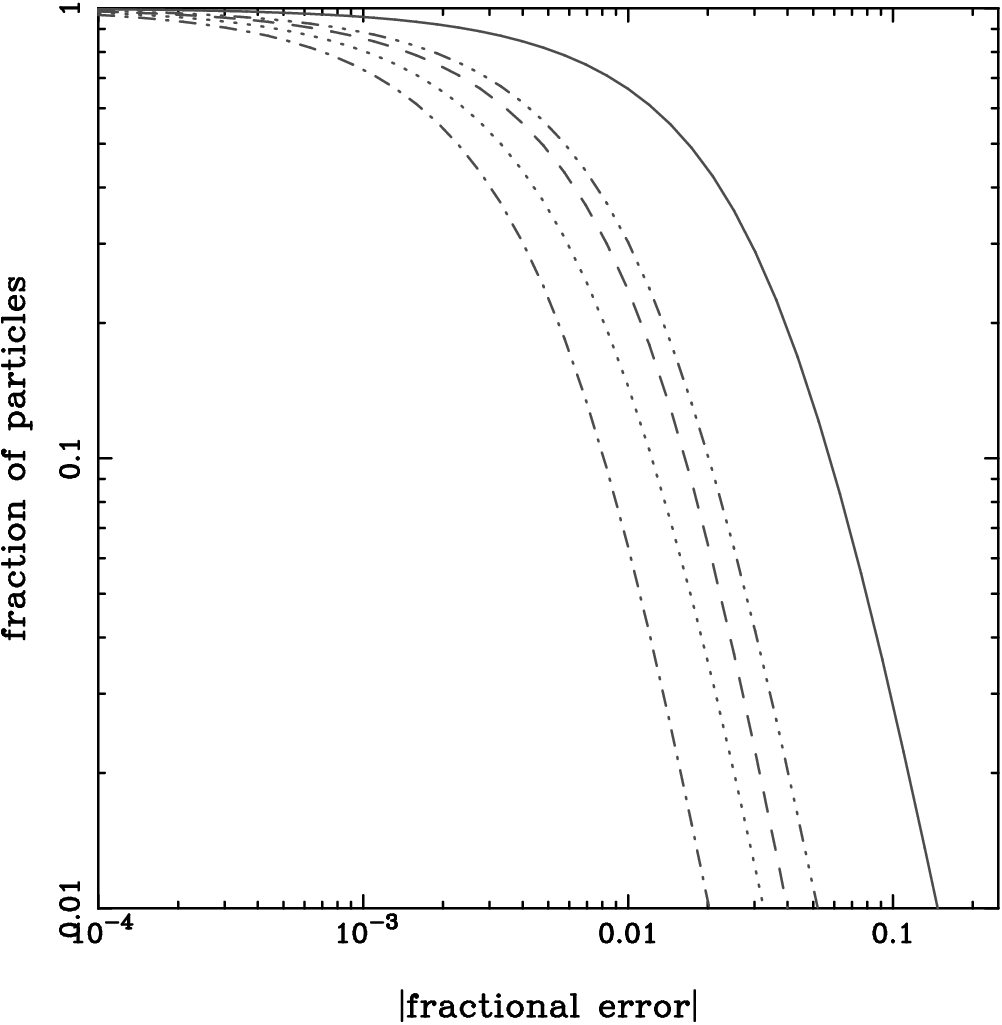}
\epsfysize=2.6truein\epsfbox[10 25 521 507]{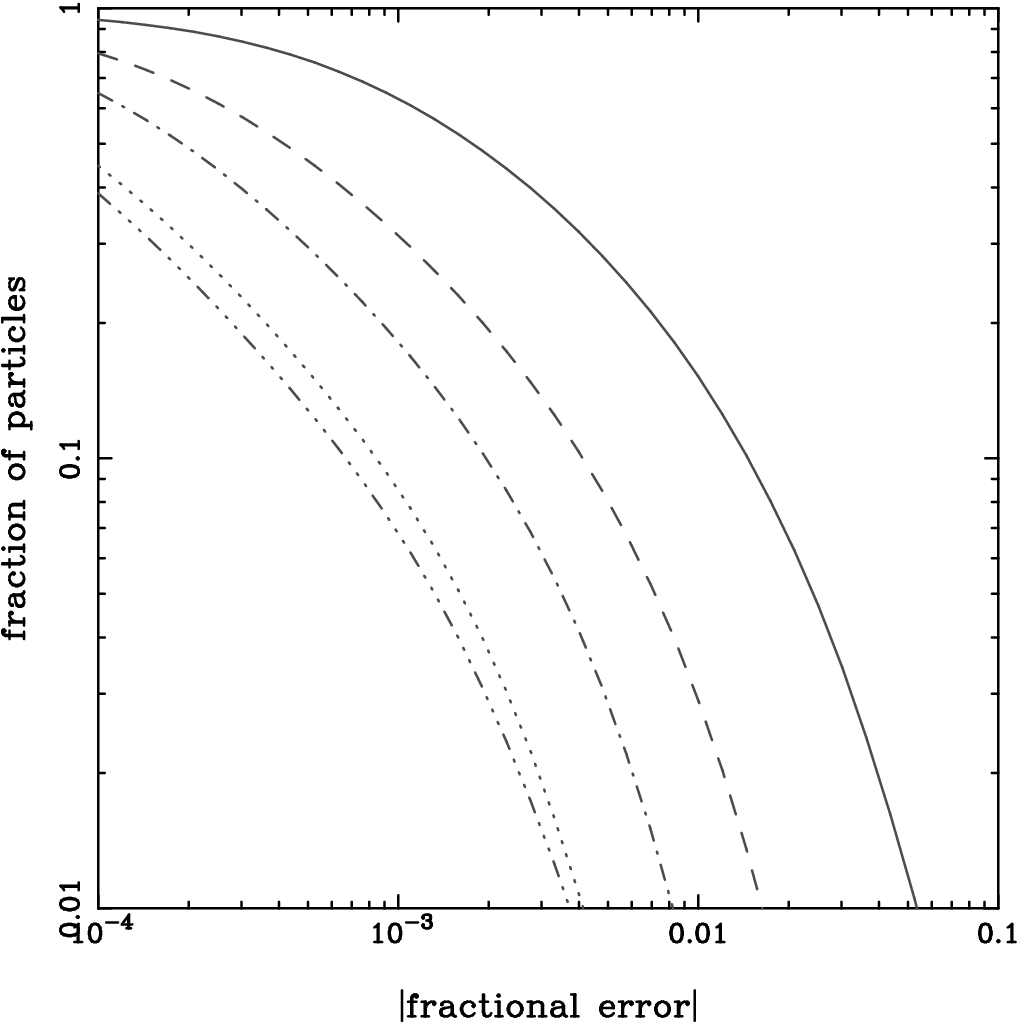}
\caption{Variation of error with $r_s$.  The left panel shows the
fractional error in TreePM force for a uniform distribution of
particles and the right panel shows the same for a clustered
distribution of particles.  We used $\theta_c=0.5$ and $r_{cut}=6r_s$
for these figures.
Different line styles are used for different values of $r_s$.  
Solid curve is for $r_s = 0.5L$, dashed curve for $r_s = 0.75L$,
dot-dashed curve for  $r_s = L$, dotted curve for  $r_s = 1.5L$ and
dot-dot-dashed curve is for  $r_s = 2L$.
The error decreases monotonically with increasing
$r_s$ for the clustered distribution.  The behaviour is different for
the unclustered distribution where the error decreases at first but
then increases again.  See text for an explanation.}
\label{fig_rs}
\end{figure}

Lastly, we look at variation of errors with $r_s$.  
Figure~{\ref{fig_rs}} shows the distribution of errors for difference
values of $r_s$.   
We fixed the values of other parameters to $r_{cut}=6r_s$ and
$\theta_c=0.5$.
Left panel is for the unclustered distribution and the right panel is
for clustered distribution of particles.  

Errors for the clustered distribution increase monotonically as $r_s$
is decreased.
This is to be expected from our analysis of errors in the long range
force.
The fractional error in the TreePM force for $99\%$ of the particles
for $r_s=L$ is less than $0.8\%$. 
Variation of errors with $r_s$ for the unclustered distribution is not
monotonic.
Errors decrease till $r_s$ increases to unity, and then increase
again.  
This happens because the small force on each particle for a uniform
distribution again comes into play.
For larger $r_s$, the force due to individual cells is large and
numerical error in adding all the large contributions to get the small
net force is also large.
The fractional error in the TreePM force for $99\%$ of the particles
for $r_s=L$ is less than $1.5\%$, thus the variation in error from
unclustered to clustered situation is small compared to that in tree
codes. 
From this figure, we conclude that $r_s=L$ is the optimum choice for
the TreePM code. 

We can summarise the results of this study of errors as follows:
Errors for 
$99\%$ of particles are below $1.5\%$ for unclustered distribution and
below $0.8\%$ for clustered distribution for $r_s=L$, $r_{cut} \geq
5r_s$ and $\theta_c \leq 0.5$.  The best way to reduce errors is to
use a larger $r_s$ as varying the other two parameters leads to large
differences between the clustered and the unclustered cases.

\section{CPU time requirements}

In this section we will list the CPU time required for one time step
of the TreePM code for a simulation of $2^{11}$ particles.  We will
also study variation of this as we vary the parameters of the TreePM
method.

The computer on which these timings were obtained is powered by a
$1.6$GHz Pentium~4 CPU.  Programs were compiled using the Intel
Fortran compiler (version 5.0) and double precision numbers were used
throughout.  Programs are written in Fortran 90/95.

The recommended configuration of the TreePM code uses $r_s = L$,
$\theta_c = 0.5$ and $r_{cut} = 5 r_s$.  In this case one time step
takes $740$~seconds. Times taken for building the tree,
calculating the long range force, calculating the short range force and 
moving particles are, for one time step, equal to 6.38 seconds, 26.70 
seconds, 711.51 seconds and 0.25 seconds respectively. 
This implies that the time taken for each particle per step is $0.35$ms.
The quoted numbers are for an unclustered uniform distribution of
$128^3$ particles.

\begin{figure}
\epsfxsize=5.4truein\epsfbox[30 20 513 521]{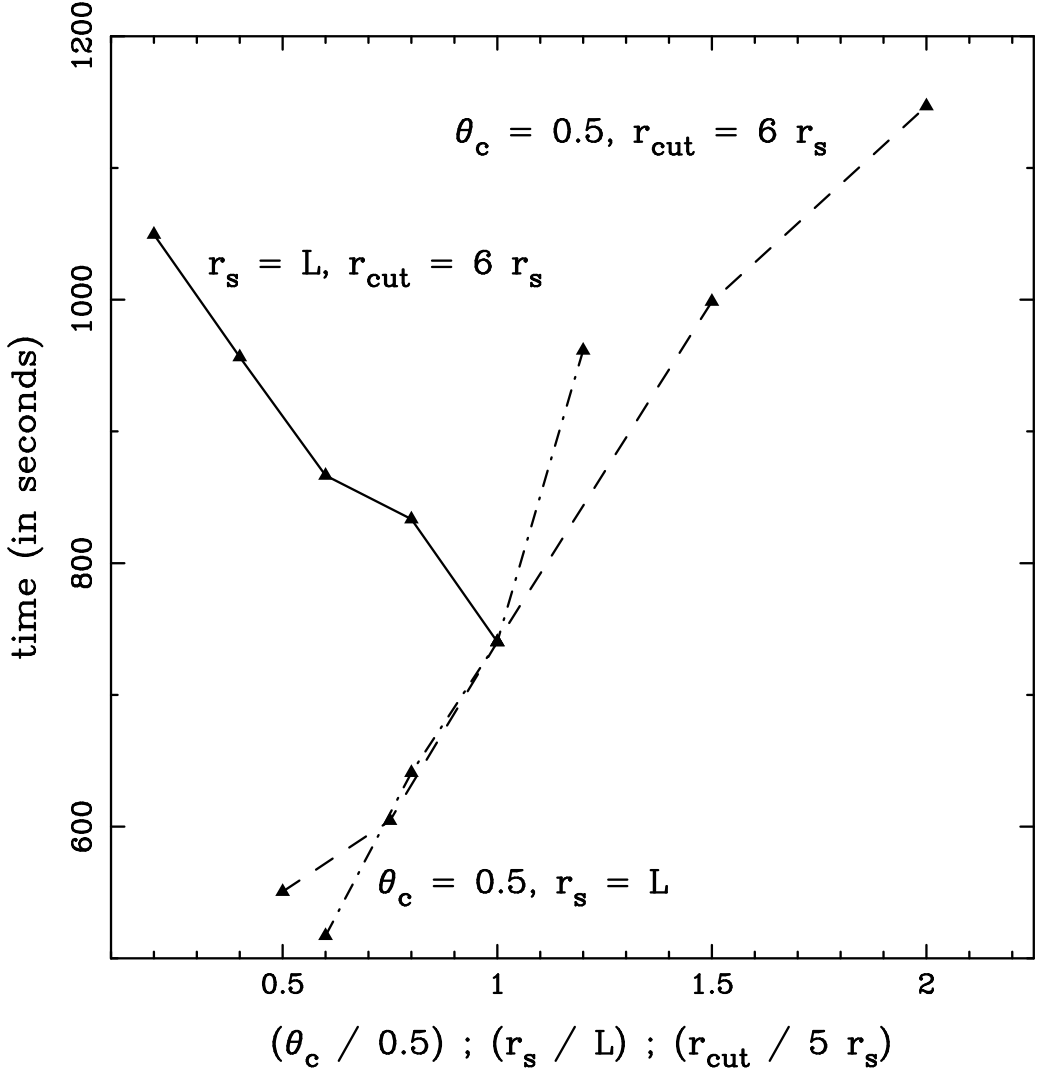}
\caption{Variation of CPU time required per time step is shown here as a
function of changing parameters.  The thick line shows the variation
in CPU time with $\theta_c/0.5$.  Points, from left to right, along
this curve are for $\theta_c=0.1, 0.2, 0.3, 0.4 $ and $0.5$.  Values
of the other parameters were fixed to $r_s=L$ and $r_{cut}=6 r_s$.
The dashed line shows the variation with respect to $r_s/L$ at
$r_{cut} =6r_s$ and $\theta_c=0.5$.  The dot-dashed line shows the
variation with $r_{cut}/5r_s$ with $\theta_c=0.5$ and $r_s = L$.}
\label{fig_timing}
\end{figure}

Time required for one time step of the TreePM code varies with the
choice of parameters $r_{cut}$, $r_s$ and $\theta_c$.  We expect the
time taken to increase with increasing $r_s$ and $r_{cut}$, and
decrease with increasing $\theta_c$.  The behaviour for these cases is
shown in fig.~{\ref{fig_timing}}.  It is seen that the time taken falls
sharply with increasing $\theta_c$, whereas errors are insensitive to
the choice as long as $\theta_c \leq 0.5$.  Therefore it makes sense
to use $\theta_c=0.5$.  Time taken rises sharply with increasing
$r_{cut}$.  Again, by comparing with the variation of errors, we
recommend $r_{cut} \approx 4.5 r_s$.  Lastly, time taken increases
rapidly as we increase $r_s$.  We
recommend $r_s=L$ as the error is minimum for this choice for the
unclustered distribution.

\section{Integrating the equation of motion}

Our discussion so far has dealt only with evaluation of force.  This
is the main focus of this paper as the key difference between the
TreePM and other methods is in the scheme used for evaluation of
force.  However for the sake of completeness, we give here details of
integration of equations of motion.  The usual form of the equation of
motion is:
\begin{eqnarray}
\ddot{\mathbf x} &+& 2 \frac{\dot{a}}{a} \dot{\mathbf x} = -
\frac{1}{a^2} \nabla\phi
\nonumber\\
\nabla^2\phi &=& 4 \pi G a^2 \left(\rho - {\bar\rho} \right)
\label{eqnmot}
\end{eqnarray}
Here ${\mathbf x}$ is the comoving coordinate, $a$ is the scale
factor, $\rho$ is the density and ${\bar\rho}$ is the average
density of the universe.  Dot represents differentiation with respect
to time.
We can recast these equations in the following form:
\begin{eqnarray}
{\mathbf x}'' &+&  \frac{2}{a}\left\{1 +
\left(\frac{1}{4}\right)\frac{2\Omega_\Lambda a^3 - \Omega_0}{\Omega_\Lambda a^3 +
\Omega_0 + a \left(1 - \Omega_0 - \Omega_\Lambda\right)} \right\}
{\mathbf x}' 
\nonumber\\
&=& - \left(\frac{3\Omega_0}{2 a^2}\right) \frac{1}{\Omega_\Lambda a^3 +
\Omega_0 + a \left(1 - \Omega_0 - \Omega_\Lambda\right)} \nabla\psi
\nonumber\\
\nabla^2\psi &=&  \delta = \frac{\rho}{\bar\rho} - 1 
\label{eqnmot_treepm}
\end{eqnarray}
Here prime denotes differentiation with respect to the scale factor.
$\Omega_0$ is the density parameter of non-relativistic matter,
$\Omega_\Lambda$ is the density parameter of the Cosmological constant
and $\delta$ is the density contrast.
We use a modified gravitational potential $\psi$.
The equation of motion contains a velocity dependent term and hence we
cannot use the usual leap-frog method.  
We recast the leap-frog method so that velocities and positions are
defined at the same instant \citep{leap_frog}.  
We solve the equation for velocity iteratively.
Time step is chosen to be a small fraction of the smallest dynamical
time in the system at any given stage. 
The fraction to be chosen is fixed by doing some tests.  

\begin{figure}
\epsfysize=2.6truein\epsfbox[31 28 524 506]{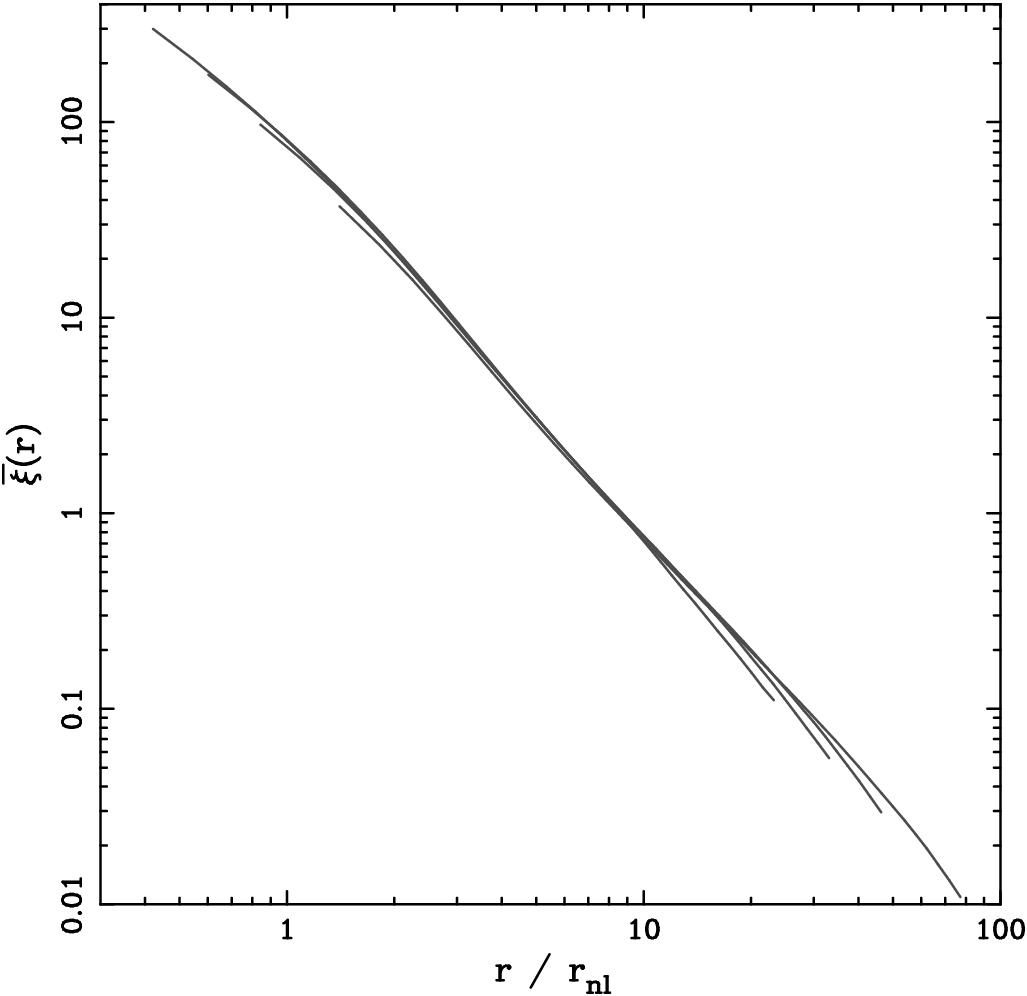}
\epsfysize=2.6truein\epsfbox[10 28 524 506]{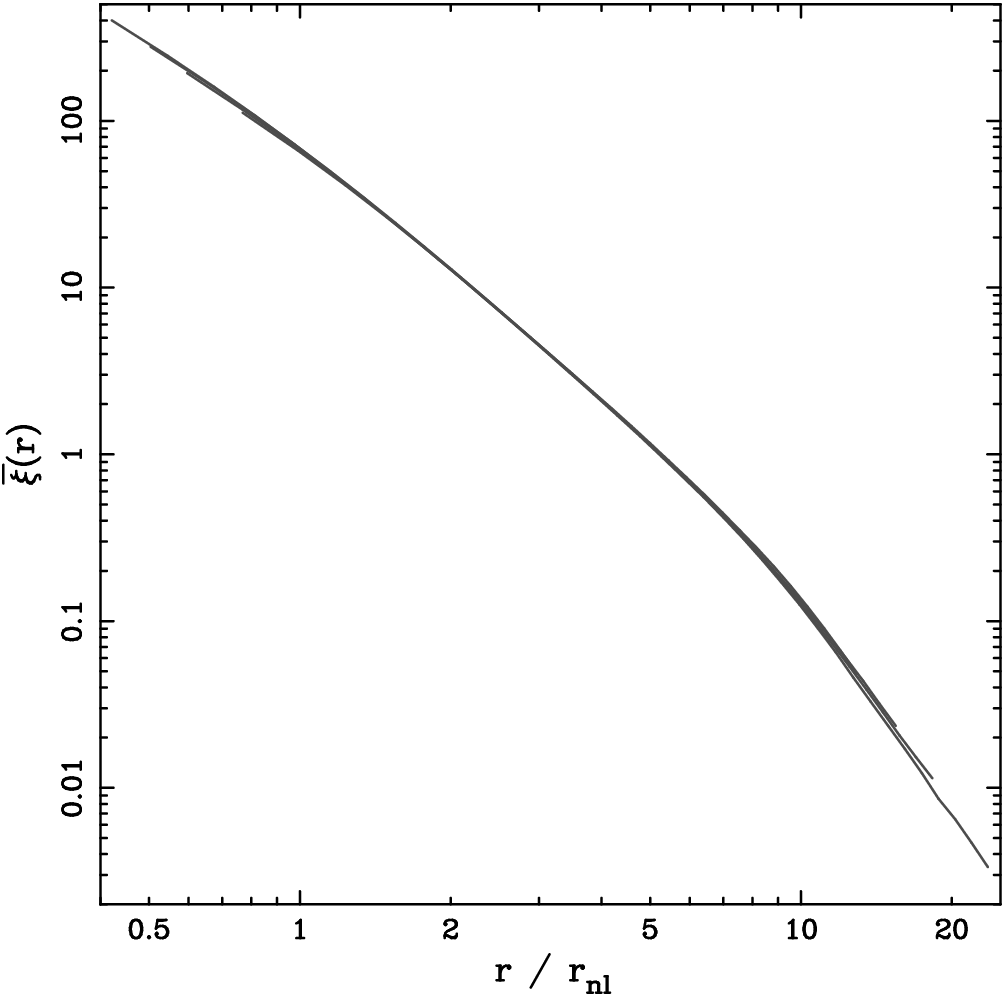}
\caption{This figure shows $\bar\xi$ as a function of
$r/r_{nl}(t)$ for several epochs.  Here $r_{nl}(t)$ is the scale which
is going non-linear at time $t$ and it varies in proportion with
$a^{2/\left(n+3\right)}$ in the Einstein-deSitter model.  $n$ is the
index of the power spectrum, we have used $n=-1$ for the left panel
and $n=1$ for the right panel.  We have only plotted $\bar\xi$ at
scales more than four times larger than the artificial softening
length used in the simulation.}
\label{fig_scale_inv}
\end{figure}

One of the tests we did was to check for scale invariance in evolution
of power law spectra.  A simulation is repeated with different choices
of time step until we find the largest time step for which we can
reach the highly non-linear regime ($\bar\xi > 100$) and retain scale
invariance.  We then use a time step that is half of this largest time
step.  Figure~{\ref{fig_scale_inv}} shows $\bar\xi$ as a function of
$r/r_{nl}(t)$ for several epochs.  Here $r_{nl}(t)$ is the scale which
is going non-linear at time $t$ and it varies in proportion with
$a^{2/\left(n+3\right)}$ in the Einstein-deSitter model.  $n$ is the
index of the power spectrum.  We have used $n=-1$ for the left panel
and $n=1$ for the right panel.  We have only plotted $\bar\xi$ at
scales more than four times larger than the artificial softening
length used in the simulation.  Scale invariance holds for both the
spectra and we are able to probe the highly non-linear regime with
this code. 

\section{Discussion}

In this paper we have presented a detailed study of performance
characteristics of the TreePM method.  We have analysed different
sources of error and suggested remedies for the main source of
errors.  The analysis of errors in realistic situations shows that
the TreePM method performs very well and gives acceptably small
errors.

This method compares favourably with other comparable methods such as
implementations of the tree code like GADGET \citep{gadget} and hybrid
methods such as the TPM \citep{tpm,tpm2}.  From the numbers available
in these papers, we find that the errors in the recommended
configuration of the TreePM method are comparable with those in GADGET
and lower than those in the TPM method.  In terms of CPU time taken
per step per particle, we again find that the numbers are comparable.
Of course, it is not possible to make a detailed comparison of this
quantity as the whole approach is different.  E.g., we do not use
multiple time steps whereas GADGET relies heavily on these to optimise
the speed.  On the other hand GADGET and TreePM have a uniform force
resolution whereas TPM does not and hence the time taken is likely to
vary more strongly with the amplitude of clustering for the TPM code
as compared to the other two.

Splitting of force into a short range and a long range part has a
useful implication in terms of parallel implementation.  Unlike the
tree methods, the TreePM method involves a considerably smaller
overhead in terms of inter-process communication and hence is likely
to score over other methods in terms of scaling for a very large
number of CPUs.  
In summary we can state that TreePM is a competitive method for doing
Cosmological N-Body simulations.  Explicit use of three parameters
gives users control over errors and CPU time required.

\section*{Acknowledgements}

A part of the work reported here was done using the Beowulf at
the Harish-Chandra Research Institute (http://cluster.mri.ernet.in).

\end{document}